# On the Generator of Lorentz Boost

**Zhi-Yong Wang**, **Cai-Dong Xiong**

*School of Physical Electronics, University of Electronic Science and Technology of China, Chengdu 610054*

E-mail: zywang@uestc.edu.cn

Traditionally, the theory related to the spatial angular momentum has been studied completely, while the investigation in the generator of Lorentz boost is inadequate. In this paper we show that the generator of Lorentz boost has a nontrivial physical significance, it endows a charged system with an electric moment, and has an important significance for the electrical manipulations of electron spin in spintronics. An alternative treatment and interpretation for the traditional Darwin term and spin-orbit coupling are given.

**PACS** numbers**:** 03.65.Ta, 71.15.Rf

## 1. Introduction

As well known, a four-dimensional (4D) angular-momentum tensor is the generator of a proper Lorentz transformation, where the purely spatial component, as the generator of a 3D spatial rotation, is the spatial angular momentum. The spatial angular momentum may provide a charged system with a magnetic moment, and its theory has been developed well. In contrary to which, the generator of Lorentz boost, as the space-time component of the 4D angular-momentum tensor, contains time coordinate that has been a controversial issue in quantum mechanics. As a result, the physical meanings of the generator of Lorentz boost seem to be vague. For example, the conservation of the generator of Lorentz boost just implies that the velocity of the relativistic inertia-center of the system is a constant [1-2].

Nowadays, nanoelectronics is progressing rapidly [3-6]. In this paper, we will show that the generator of Lorentz boost may endow a charged system with an electric moment, and then has theoretical and applied interests in nanoelectronics. For example, the potential technology significance of an interaction related to the generators of Lorentz boost lies in that, it may offer purely electrical ways to control and manipulation of electron spin in spintronics.

In the following we apply the natural units of measurement ($\hbar = c = 1$), let the metric tensor $g^{\mu\nu} = \text{diag}(1, -1, -1, -1)$ and denote 4D space-time coordinates as $x^\mu = (\boldsymbol{x}, t)$. Furthermore, repeated indices must be summed according to the Einstein rule.

## 2. The Generator of Lorentz boost and electric moment

In special relativity, the concept of 3D orbital angular momentum $\boldsymbol{L} = \boldsymbol{x} \times \boldsymbol{p}$ is extended to that of 4D orbital-angular-momentum tensor $L^{\mu\nu} = x^\mu p^\nu - x^\nu p^\mu$, where the space-time component corresponds to the generator of Lorentz boost. Correspondingly, the relation between the 3D angular momentum and magnetic moment can be extended to the 4D case. As for a 4D electromagnetic potential $A_\mu(x)$, in an analogous manner (see for example, Ref. [7]), one has

$$A^\mu(\boldsymbol{x},t) = \frac{1}{4\pi}\int \frac{J^\mu(\boldsymbol{x}',t')}{|\boldsymbol{x}-\boldsymbol{x}'|} \mathrm{d}^3\boldsymbol{x}' = \frac{1}{|\boldsymbol{x}|}\int J^\mu(\boldsymbol{x}',t')\mathrm{d}^3\boldsymbol{x}' + \frac{\boldsymbol{x}'}{|\boldsymbol{x}|^3}\int J^\mu(\boldsymbol{x}',t')\boldsymbol{x}'\mathrm{d}^3\boldsymbol{x}' + \cdots \quad (1)$$

where $t' = t - |\boldsymbol{x}|/c = t - |\boldsymbol{x}|$ ($\hbar = c = 1$), $J^\mu$ is a localized divergenceless current, which permits simplification and transformation of the expansion (1). Let $f(x')$ and $g(x')$ be well-behaved functions of $x'^\mu = (\boldsymbol{x}', t')$ to be chosen below

$$\int (fJ \cdot \Box g + gJ \cdot \Box f)\mathrm{d}^4 x = 0 \qquad (\Box \cdot J = \partial^\mu J_\mu = 0), \qquad (2)$$

where $\Box$ denotes the 4D gradient operator. Similar to Ref. [7], let $f = x_\mu$ and $g = x_\nu$, one has

$$\int (x_\mu J_\nu + x_\nu J_\mu)\mathrm{d}^4 x = 0, \qquad (3)$$

and then

$$\boldsymbol{x} \cdot \frac{\mathrm{d}}{\mathrm{d}t'} \int x' J_\mu(x') \mathrm{d}^4 x' = -\frac{1}{2} \sum_j x_j \frac{\mathrm{d}}{\mathrm{d}t'} \int (x_i' J_\mu - x_\mu' J_j) \mathrm{d}^4 x' \qquad (4)$$

It is customary to define the 4D electromagnetic-moment-density tensor

$$m^{\mu\nu} = \frac{1}{2}[x^\mu J^\nu - x^\nu J^\mu] \qquad (5)$$

with its integral as the electromagnetic moment (not a tensor)

$$M^{\mu\nu} = \frac{1}{2}\int [x'^\mu J^\nu - x'^\nu J^\mu] \mathrm{d}^3 x' . \qquad (6)$$

Assuming that $J^\mu$ is provided by $N$ charged particles with momenta $p_n^\mu = m u_n^\mu$ and charges $e$ ($n=1,2,\ldots N$, $m$ is the rest mass, $u_n^\mu$ is the 4-velocity of the $n$-th particle), one has

$$J^\mu(x') = \sum_n e u_n^\mu(t') \delta^3(\boldsymbol{x}-\boldsymbol{x}_n') \frac{\mathrm{d}\tau}{\mathrm{d}\tau'} \qquad (7)$$

where $\tau$, $\tau'$ are the proper times. Taking $t = t'$, and let $m'$ denote the relativistic mass, one has

$$M^{\mu\nu} = \frac{e}{2m'} L^{\mu\nu} = \frac{e}{2m} \sum_n (x_n^\mu p_n^\nu - x_n^\nu p_n^\mu) \frac{\mathrm{d}\tau}{\mathrm{d}t} \qquad (8)$$

where $L^{\mu\nu} = \sum_n (x_n^\mu p_n^\nu - x_n^\nu p_n^\mu)$ is the 4D total orbital-angular-momentum tensor.

Therefore, in the multipole expansion of the 4D electromagnetic potential of a charged system, the 4D angular-momentum tensors contribute the magnetic and electric moments to the system. In other words, a 3D spatial angular momentum may result in a magnetic moment, while the generator of Lorentz boost may bring an electric moment. For these a heuristic understanding or an intuitive physical picture can be obtained from Ref. [8].

## 3. 4D spin-orbit tensor coupling of the electron

The generator of Lorentz boost discussed above corresponds to the space-time component of a 4D orbital-angular-momentum tensor. Now, in the level of quantum mechanics, we will turn to the infinitesimal generator of Lorentz boost, which corresponds to the space-time component of a 4D spin tensor. Consider an infinitesimal proper Lorentz transformation that changes 4D coordinates according to $x^\mu \to x'^\mu = (g^{\mu\nu} + \varepsilon^{\mu\nu})x_\nu$ ($\mu, \nu = 0, 1, 2, 3$), where $\varepsilon^{\mu\nu}$ is an infinitesimal antisymmetric tensor. Owing to the Lorentz symmetry, the Dirac field $\psi(x) = \psi(\boldsymbol{x}, t)$ of the electron is transformed as

$$\psi(x) \to \psi'(x') = (1 - \frac{\mathrm{i}}{2}\varepsilon_{\mu\nu}S^{\mu\nu})\psi(x) \qquad (9)$$

where the 4D spin tensor $S^{\mu\nu} = -S^{\nu\mu}$ of the electron is called the infinitesimal generator of Lorentz transformations of $\psi(x)$, its space component corresponds to the infinitesimal generator of spatial rotations, while the component with mixed spatial-temporal indices represents the infinitesimal generator of Lorentz boosts. Let $\hat{p}_\mu = \mathrm{i}\partial/\partial x^\mu \equiv \mathrm{i}\partial_\mu$ denote the 4D momentum operator, under Eq. (9) the total variation of $\psi(x)$ is $\delta\psi = \psi'(x) - \psi(x) = -\mathrm{i}\varepsilon_{\mu\nu}\hat{J}^{\mu\nu}\psi(x)/2$, where $\hat{J}^{\mu\nu} = x^\mu\hat{p}^\nu - x^\nu\hat{p}^\mu + S^{\mu\nu}$ is the 4D total-angular-momentum tensor.

Let $A^\mu = (A^0, \boldsymbol{A})$ ($\mu = 0, 1, 2, 3$) be the 4D potential of an external electromagnetic field ($\boldsymbol{A}$ is the vector potential and $A^0 = \Phi$ the scalar potential), $e$ the unit charge, $m$ the electron mass, and $D_\mu \equiv \partial_\mu + \mathrm{i}eA_\mu$ denote the gauge covariant derivative, the Dirac equation of the electron in $A^\mu$ is

$$(\mathrm{i}\gamma^\mu D_\mu - m)\psi(x) = 0 \qquad (10)$$

where the four 4×4 Dirac matrices $\gamma^\mu$ satisfy the algebra $\gamma^\mu\gamma^\nu + \gamma^\nu\gamma^\mu = 2g^{\mu\nu}$, in terms of them the 4D spin tensor can be written as $S^{\mu\nu} = \mathrm{i}(\gamma^\mu\gamma^\nu - \gamma^\nu\gamma^\mu)/4$. Let $\varepsilon^{ijk}$ denotes the totally anti-symmetric tensor with $\varepsilon^{123} = 1$ ($i, j, k = 1, 2, 3$), one can show that $\boldsymbol{\Sigma} = (\Sigma_1, \Sigma_2, \Sigma_3)$ with $\Sigma_i = \varepsilon_{ijk}S^{jk}/2$ is the usual spin matrix (i.e. the infinitesimal generator of 3D spatial rotation). As an analogy we call $\boldsymbol{K} \equiv (S^{01}, S^{02}, S^{03})$ as spin-like matrix (corresponds to the infinitesimal generator of Lorentz boost). In terms of the Pauli's matrix vector $\boldsymbol{\sigma}=(\sigma_1,\sigma_2,\sigma_3)$, the spin matrix $\boldsymbol{\Sigma}$ and spin-like matrix $\boldsymbol{K}$ can be expressed as, respectively

$$\boldsymbol{\Sigma}=\frac{1}{2}\begin{pmatrix}\boldsymbol{\sigma}&0\\0&\boldsymbol{\sigma}\end{pmatrix},\quad \boldsymbol{K}=\frac{i}{2}\begin{pmatrix}0&\boldsymbol{\sigma}\\\boldsymbol{\sigma}&0\end{pmatrix} \qquad (11)$$

From Eq. (10) one can give a second-order one [9]:

$$[D^\mu D_\mu + m^2 + eS^{\mu\nu}F_{\mu\nu}]\psi(x)=0 \qquad (12)$$

where $F_{\mu\nu}=\partial_\mu A_\nu - \partial_\nu A_\mu$ is the electromagnetic field tensor. By the proper time $\tau=\sqrt{x^\mu x_\mu}$ one has $\partial_\mu = \partial/\partial x^\mu = x_\mu \partial/\tau\partial\tau$. Using $\psi = i\gamma^\mu D_\mu \psi/m$ (see Eq. (10)), $a^\mu \gamma_\mu b^\nu \gamma_\nu = a^\mu b_\mu - 2ia^\mu b^\nu S_{\mu\nu}$ and the Lorentz gauge condition $\partial_\mu A^\mu = 0$, Eq. (12) becomes ($\mu,\nu,\alpha,\beta = 0,1,2,3$)

$$[D^\mu D_\mu + m^2 + \frac{e}{m}\gamma_\mu(\partial_\nu A^\mu)D^\nu - \frac{e}{m}\kappa S_{\alpha\beta}\hat{L}_A^{\alpha\beta}]\psi(x)=0 \qquad (13)$$

where $\hat{L}_A^{\alpha\beta} \equiv x^\alpha iD^\beta - x^\beta iD^\alpha$ is the 4D orbit-angular-momentum tensor with the 4D momentum operator $\hat{p}_\mu = i\partial_\mu$ being replaced by $iD_\mu \equiv i[\partial_\mu + ieA_\mu]$, $\kappa \equiv \gamma^\mu \partial A_\mu/\tau\partial\tau$ is a matrix parameter. Eq. (13) is for the first time obtained by us. The term related to $S_{\alpha\beta}\hat{L}_A^{\alpha\beta}$ is called as the 4D spin-orbit tensor coupling, which describes the coupling between the 4D spin tensor and the 4D orbit-angular-momentum tensor, and includes not only the usual 3D spatial spin-orbit coupling, but also the interaction related to the infinitesimal generators of Lorentz boost.

On other hand, let $\psi = \psi' \exp(-imt)$, in the approximation $|i\partial\psi'/\partial t|$, $|e\Phi\psi'| \ll m|\psi'|$ and $|i\partial\Phi/\partial t| \ll m|\Phi|$, in term of the electric and magnetic fields $\boldsymbol{E}=-\nabla\Phi-\partial_t\boldsymbol{A}$, $\boldsymbol{B}=\nabla\times\boldsymbol{A}$, Eq. (12) (or Eq. (13)) becomes a semi-nonrelativistic approximation form

$$i\frac{\partial}{\partial t}\psi' = [\frac{(\hat{\boldsymbol{p}}-e\boldsymbol{A})^2}{2m} + e\Phi - \frac{e}{m}\boldsymbol{\Sigma}\cdot\boldsymbol{B} + \frac{e}{m}\boldsymbol{K}\cdot\boldsymbol{E}]\psi' \qquad (14)$$

Eq. (14) may be found in other literatures, however, here we look at Eq. (14) from a different point of view. That is, Eq. (14) shows that the electron has a magnetic moment $e\boldsymbol{\Sigma}/m$ and an electric moment $e\boldsymbol{K}/m$. We refer to $e\boldsymbol{\Sigma}/m$ as *intrinsic* magnetic moment while $e\boldsymbol{K}/m$ as *induced* electric moment. Therefore, in terms of the concept of induced electric moment, an alternative treatment and interpretation for the Darwin term and the 3D spatial spin-orbit coupling are given.

In a word, the spin matrix $\boldsymbol{\Sigma}$, as the infinitesimal generator of 3D spatial rotation, provides the electron with an intrinsic magnetic moment, while the spin-like matrix $\boldsymbol{K}$, as infinitesimal generator of Lorentz boost, provides the electron with an induced electric moment. Eqs. (13)-(14) again show that the generator of Lorentz boost has a nontrivial physical significance. However, let us emphasize that the so-called intrinsic or permanent dipole electric moment of the electron that people are searching for nowadays (in order to provide an interpretation for the violation of time reversal invariance) is conceptually distinct from the induced electric moment discussed here. The interaction between the induced electric moment and electric fields does not break time reversal symmetry.

## 4. An example of application

Conventionally, electronics only sensitive to electron's charge, spin degrees of freedom are ignored. Spintronics, a new research field develop in recent years, is based on the up (↑) or down (↓) spin of carriers rather than on electrons or holes as in traditional semiconductor electronics, and involves the study of active control and manipulation of spin degrees of freedom in solid-state systems [10]. It is crucial for spintronics device applications to manipulate electron spins by purely electric means [11-17].

Now we give an illustration for purely electrostatic manipulations of electron spin via the degrees of freedom related to the infinitesimal generator of Lorentz boost. Let $\psi' = (\varphi' \; \chi')^T$, where the superscript T denotes matrix transpose. Eq. (14) becomes

$$\begin{cases} i\dfrac{\partial}{\partial t}\varphi' = [\dfrac{(\hat{\boldsymbol{p}}-e\boldsymbol{A})^2}{2m} + e\Phi - \dfrac{e}{2m}\boldsymbol{\sigma}\cdot\boldsymbol{B}]\varphi' + i\dfrac{e}{2m}\boldsymbol{\sigma}\cdot\boldsymbol{E}\chi' \\ i\dfrac{\partial}{\partial t}\chi' = [\dfrac{(\hat{\boldsymbol{p}}-e\boldsymbol{A})^2}{2m} + e\Phi - \dfrac{e}{2m}\boldsymbol{\sigma}\cdot\boldsymbol{B}]\chi' + i\dfrac{e}{2m}\boldsymbol{\sigma}\cdot\boldsymbol{E}\varphi' \end{cases} \quad (15)$$

Let $\psi'_+ = \varphi' + \chi'$ and $\psi'_- = \varphi' - \chi'$, Eq. (15) becomes

$$i\dfrac{\partial}{\partial t}\begin{pmatrix}\psi'_+\\ \psi'_-\end{pmatrix} = \hat{H}\begin{pmatrix}\psi'_+\\ \psi'_-\end{pmatrix} = \begin{pmatrix}\hat{H}_0+\hat{H}_1 & 0 \\ 0 & \hat{H}_0-\hat{H}_1\end{pmatrix}\begin{pmatrix}\psi'_+\\ \psi'_-\end{pmatrix} \quad (16)$$

where

$$\hat{H}_0 = \dfrac{(\hat{\boldsymbol{p}}-e\boldsymbol{A})^2}{2m} + e\Phi - \dfrac{e}{2m}\boldsymbol{\sigma}\cdot\boldsymbol{B}, \quad \hat{H}_1 = i\dfrac{e}{2m}\boldsymbol{\sigma}\cdot\boldsymbol{E} \quad (17)$$

Now, in order to emphasize the interactions related to the infinitesimal generators of Lorentz boost, one assumes $\boldsymbol{B}=0$ and $\partial \boldsymbol{E}/\partial x^\mu = 0$ ($\mu=0,1,2,3$), then $[\hat{H}_0, \hat{H}_1]=0$, $\hat{H}_0$ and $\hat{H}_1$ have common eigenstates. In this case, one has

$$\hat{H}_0\begin{pmatrix}\psi'_+\\ \psi'_-\end{pmatrix} = \varepsilon_0\begin{pmatrix}\psi'_+\\ \psi'_-\end{pmatrix}, \quad \hat{H}_1\begin{pmatrix}\psi'_+\\ \psi'_-\end{pmatrix} = \varepsilon_1\begin{pmatrix}\psi'_+\\ \psi'_-\end{pmatrix} \quad (18)$$

where, obviously, $\varepsilon_0$ and $\varepsilon_1$ are real eigenvalues. Then Eq. (16) becomes

$$\hat{H}\begin{pmatrix}\psi'_+\\ \psi'_-\end{pmatrix} = \begin{pmatrix}\varepsilon_0-\varepsilon_1 & 0 \\ 0 & \varepsilon_0+\varepsilon_1\end{pmatrix}\begin{pmatrix}\psi'_+\\ \psi'_-\end{pmatrix} \quad (19)$$

From Eqs. (16)-(19), one can arrival at the conclusions as follows:

1) As $\boldsymbol{E}=0$, $\psi'_+$ and $\psi'_-$ are two degeneration states of $\hat{H}$ with the same eigenvalue $\varepsilon_0$;

2) As $\boldsymbol{E}\neq 0$, the eigenvalue $(\varepsilon_0-\varepsilon_1)$ associated with the eigenstate $\psi'_+$ is no longer equal to the eigenvalue $(\varepsilon_0+\varepsilon_1)$ associated with the eigenstate $\psi'_-$, and then the original two-fold degeneracy is broken by the presence of the electrostatic field $\boldsymbol{E}$. As a result, an energy-level splitting of $2\varepsilon_1$ between the eigenstates $\psi'_+$ and $\psi'_-$ is produced.

As well known, a two energy-level system is often taken as an information-carrier that is indispensable for information processing. For example, Spintronics is based on an energy-level splitting between the spin-up and spin-down states of the electron in an external field. Eq. (19) shows that, via the interaction associated with the infinitesimal generator of Lorentz boost, one can implement an electrostatic manipulation of electron spin.

## 5. Conclusions

Owing to the fact that space coordinates play the role of dynamical variables while time coordinate just plays the role of a parameter, time in quantum mechanics has been a controversial issue since the advent of quantum theory [18-19]. Correspondingly, the generator of Lorentz boost contains time coordinate such that its physical meaning seems to be ambiguous or trivial. As a result, traditionally the theoretical and application interests are mainly focused on the 3D spatial angular momentum rather than the 4D angular-momentum-tensor itself. In the paper, we have shown that a charged system with nonzero generator of Lorentz boost may have an electric moment, just as that a charged system with nonzero generator of spatial rotation may have a magnetic moment. Starting from the Dirac equation of the electron in external fields, one can generally obtain the coupling interaction between the 4D spin tensor and the 4D orbit-angular-momentum tensor, which includes the interaction related to the intrinsic magnetic moment of the electron and that associated with the infinitesimal generator of Lorentz boost, and they together offer all possible electromagnetic ways to control and manipulation of electron spin in spintronics.

**Acknowledgements**. We would like to acknowledge the financial support of the Doctoral

Program Foundation of Institution of Higher Education of China (Grant No. 20050614022).## References

[1] Ticciati R 1999 *Quantum Field Theory for Mathematicians* (England: Cambridge University Press) pp65-67.

[2] Landau L D and Lifshitz E M 1999 *The Classical Theory of Fields* (Singapore: Beijing World Publishing Corporation) pp40-42.

[3] Chen Z Q ,Chen H ,Cheng N P and Zheng R L 2002 *Acta Phys .Sin*. **51** 649 (in Chinese)

[4] Zheng R L , Zhang C L and Lu J 2003 *Acta Phys . Sin.* **52** 2284 (in Chinese)

[5] Zheng R L , Zhang C L and Chen Z Q 2005 *Acta Phys . Sin.* **54** 886 (in Chinese)

[6] Zheng R L and Wen G Z 2006 *Acta Phys . Sin.* **55** 791 (in Chinese)

[7] Jackson J D 1975 *Classical Electrodynamics* (England: John Wiley & Sons. Inc.) p180.

[8] Grosse C 1984 *Am. J. Phys*. **52** 125.

[9] Scharf G 1995 *Finite Quantum Electrodynamics* (Berlin: Springer-Verlag) p 64

[10] Zutic I, Fabian J and Sarma S D 2004 *Rev. Mod. Phys*.**76** 323.

[11] Congjun W and Shou-Cheng Z 2004 *Phys. Rev. Lett.* **93** 036403.

[12] Salis G, KatoY, Ensslin K *et al* 2001 *Nature* **414** 619.

[13] Kato Y, Myers R C, Gossard A C *et al* 2004 *Nature* **427** 50.

[14] Flatté M E 2004 *Nature* **427** 21.

[15] Wolf S A 2001 *Science* **294** 1488.

[16] Rashba E I and Efros Al L 2003 *Phys. Rev. Lett*. **91** 126405.

[17] Jedema F J, Heersche H B, Filip A T *et al* 2002 *Nature* **418** 713.

[18] Wang Z Y, Chen B and Xiong C D 2003 *J. Phys. A: Math. Gen*. **36** 5135

[19] Muga J G, Sala R and Egusguiza I L 2002 *Time in Quantum Mechanics* (Berlin: Springer) pp 279–304
9